\documentclass[runningheads]{llncs}
\usepackage[T1]{fontenc}

\usepackage{graphicx}
\usepackage{caption}
\usepackage{amsmath}
\usepackage{tikz}
\tikzset{%
  Line Width/.code={%
    \pgfpointxy{#1}{0}%
    \pgfgetlastxy\tmpx\tmpy\tikzset{line width/.expanded=\tmpx}%
  },
  person/.style={
    Line Width=1/4, draw=black, color=gray!50
  },
  highlight/.style={
    preaction={Line Width=1/4, draw=gray!50, path fading=west, fading angle=-45},
    Line Width=1/4, draw=white, , path fading=east, fading angle=-45,
}}
\tikzset{>=latex}

\usepackage{subcaption}

\begin{document}

\interlinepenalty=-1
\setlength{\textfloatsep}{5pt}

\setlength{\abovedisplayskip}{3pt}
\setlength{\belowdisplayskip}{3pt}

\title{Short Squeeze in DeFi Lending Market: Decentralization in Jeopardy?}
%
%

\author{Lioba Heimbach\and 
Eric Schertenleib
\and 
Roger Wattenhofer}
\authorrunning{L. Heimbach et al.}
%
\institute{
\email{\{hlioba,ericsch,wattenhofer\}@ethz.ch}\\
ETH Zürich, Switzerland}
\maketitle              
\begin{abstract}
Anxiety levels in the Aave community spiked in November 2022 as Avi Eisenberg performed an attack on Aave. Eisenberg attempted to short the CRV token by using funds borrowed on the protocol to artificially deflate the value of CRV. While the attack was ultimately unsuccessful, it left the Aave community scared and even raised question marks regarding the feasibility of large lending platforms under decentralized governance. 

In this work, we analyze Avi Eisenberg's actions and show how he was able to artificially lower the price of CRV by selling large quantities of borrowed CRV for stablecoins on both decentralized and centralized exchanges. Despite the failure of his attack, it still led to irretrievable debt worth more than 1.5 Mio USD at the time and, thereby, quadrupled the protocol's irretrievable debt. Furthermore, we highlight that his attack was enabled by the vast proportion of CRV available to borrow as well as Aave's lending protocol design hindering rapid intervention. We stress Eisenberg's attack exposes a predicament of large DeFi lending protocols: limit the scope or compromise on `decentralization'.

\end{abstract}

\section{Introduction}

While borrowing assets serves a whole host of purposes, perhaps none have gained such an infamous reputation as short-selling. Thereby, a market participant borrows funds but immediately upon entering the borrowing contract, sells the asset (`selling it short') in the hope of re-acquiring it (`covering the short') before the lending contract expires at a lower price. A short-seller thus profits if the asset loses value. However, short-selling naturally involves risks for the speculator -- mainly that the asset appreciates in value. In this case, the borrower must buy back the asset at a higher price than they sold it for. In dramatic cases, the borrower becomes at risk of defaulting on the loan, and the lender may demand the repayment, thus, forcing the borrower to cover the short (`short squeeze').

In traditional finance, banks or brokers provide loans that enable short-selling. The conditions of the loan are typically set by the lender or are agreed upon by the involved parties. Thus, the loan must be approved by the lender, and the lender naturally closely monitors the financial situation of the borrower and employs complex active risk management to safeguard their funds. 

In contrast, \textit{Decentralized Finance (DeFi)} aims to build a financial ecosystem that provides the financial services of the traditional financial sector without relying on a central authority or placing any trust in a counterparty. Given its role as a cornerstone of finance, it is not surprising that DeFi protocols were launched that enable borrowing and lending. One of the most well-known lending protocols is \textit{Aave}~\cite{2021aave}. On Aave's V2-version users could borrow a host of different ERC20 tokens, including the \textit{Curve DAO token (CRV)}, the utility token of the decentralized exchange Curve~\cite{2023curve}. The key to offering loans in a trustless setting is requiring users to deposit collateral of greater value than the debt taken out (\textit{over-collateralization}). Loans at risk of becoming under-collateralization are offered for liquidation at a discount to protect lenders and avoid irretrievable debt on the protocol.

In October 2022, Avi Eisenberg began publicly discussing ideas on how to attack Aave, i.e., burdening the protocol with irretrievable debt for his profit. He had previously gained a reputation by extracting more than 100 Mio USD from the DeFi protocol Mango Markets~\cite{Mango_Markets} using price manipulation -- fooling the protocol about the value of the pledged collateral~\cite{blockworks_Avi_attack_Mango}. The ideas discussed for Aave were of similar nature. Then, in November 2022, Avi Eisenberg began his attack on Aave by shorting CRV. In particular, he deposited USDC, a stablecoin\footnote{Stablecoins are cryptocurrencies that are pegged to a `stable' asset -- typically the USD. For our purposes, we can use USDT, USDC, and BUSD (Binance USD) as synonymous with USD.}, as collateral on Aave and sold CRV short. His actions led to a rapid price decrease and thereby devalued the collateral of users who had deposited CRV. These developments lead to significant anxiety in the Aave community. They feared being left with irretrievable debt, which ultimately did materialize, despite the failure of the attack.

\textbf{Our Contribution.} In this work, we dissect Avi Eisenberg's actions and use this as a case study to highlight threats to the viability of DeFi lending protocols. In particular, we observe Eisenberg selling his borrowed CRV on both decentralized and centralized exchanges for stablecoins. We further analyze Eisenberg's effect on the CRV borrowing market and show that Eisenberg selected CRV for a good reason. Aave offered significant liquidity relative to CRV's market capitalization enabling him to greatly affect the price of CRV using the borrowed assets. Finally, we discuss the aftermath of the attack. We show that a single (failed) attack quadrupled the amount of bad debt on Aave, underscoring the threat posed to the whole DeFi lending market. Further, we review the actions taken by Aave to rectify the situation and discuss the viability of `decentralized' lending platforms.

\section{Background}
\label{sec: Background}
Lenders require that the borrower puts up collateral to secure the loan. While in traditional finance, this `collateral' is at times only limited to the borrower's creditworthiness and their promise to repay (unsecured debt), this approach does not adhere to the trustless principle of decentralized finance. Furthermore, both borrowers and lenders often rely on third-party intermediaries like banks to facilitate the transaction.

DeFi lending protocols, on the other hand, seek to offer these services without a trusted third-party or prior clearance of either borrower or lender. Instead, users deposit assets that the protocol supports, such as CRV for Aave V2, thereby becoming \textit{liquidity providers}. Liquidity providers earn interest on their funds locked in the protocol. Users seeking to borrow can use the deposited funds as collateral to take out debt. The required collateral significantly exceeds their maximal debt in value. Thus, the borrower is required to \textit{over-collateralize}. This over-collateralization allows users to borrow without prior clearance.

The parameters of Aave's smart contract determine the terms of borrowing and lending. The formula for the rates is included in Appendix~\ref{app_sec:borrowingrates}. Qualitatively, it is important to note that the rates are determined by the \textit{utilization}, i.e., the fraction of deposited funds that have been borrowed, and the risk parameters set by the protocol in accordance to the perceived risk of the asset. Furthermore, note that there is a maximal borrowing rate that can be reached. In the case of CRV, the rate is limited to $307\%$. 

In the event that the collateralization of the borrower becomes insufficient, the position can be liquidated. The threshold for this is given by the \textit{health factor} $H$ reaching 1, where 
\begin{equation*}
    H = \frac{\sum _{i\in A} ( C_i \cdot l_i) }{\sum _{i\in A} D_i}.
\end{equation*}
Here, the sum running over the set assets available on the protocol $A$, $C_i$ is the collateral deposited in currency $i$, $l_i$ is the liquidation threshold for asset $i$, and $D_i$ is the debt in token $i$. Once the health factor drops below 1, its collateral is auctioned off at a discount to a liquidator who must repay the debt. In the case of CRV, Aave V2 uses a liquidation threshold of $89\%$.

To calculate the health factor and other asset price dependent quantities, lending protocols rely on \textit{price oracles}. For example, Aave V2 uses Chainlink's price oracle~\cite{2023chainlink}. Chainlink thus provides Aave with information on the price the asset is currently trading at on centralized markets. 
Finally, we note that like many DeFi protocols, Aave is governed by a \textit{decentralized autonomous organization (DAO)}. Holders of the Aave token (native token of Aave) comprise the DAO of Aave. They can vote on proposals such as changing the protocol's risk parameters. The procedure for such changes must abide to rules that are enforced by the protocol's smart contract. Therefore, such changes do take several days to get implemented.

\section{Related Work}

Early research by Bartoletti et al.~\cite{bartoletti2021sok} on DeFi lending protocols, which became widely adopted amidst the excitement of the 2020 DeFi summer, provides a systematization of knowledge regarding lending protocols as well as a formal framework to model them. Gudgeon et al.~\cite{gudgeon2020defi} further present an empirical study of interest rate rules utilized by lending protocols. 

Lending protocols are a DeFi corner stone, as Aramonte et al.~\cite{aramonte2022defi} point out they allow users to easily take on leverage and are mainly used to facilitate cryptocurrency price speculation. The central position of lending protocols in DeFi, thus, makes cryptocurrency prices increasingly sensitive as Chiu et al.~\cite{chiu2022inherent} note. Our work studies exactly how leveraged trading on lending protocols affects cryptocurrency prices by examining a case study in detail.

Heimbach et al.~\cite{heimbach2023defi} investigate the effects of the Merge on Ether rates on Aave and Compound. They discuss how the DAOs of the respective protocols took actions to prevent exorbitant rates leading to mass liquidations. In contrast to the Merge, the short-selling we investigate was not announced well in advanced, giving less time for the protocols to adapt and thus posing a different kind of challenge to the lending protocols. 

Attacks, arbitrage opportunities and trading strategies on DeFi protocols are frequent and well-studied studied in previous work~\cite{eskandari2021sok,qin2022quantifying,yaish2022blockchain,perez2021liquidations,qin2021empirical,torres2021frontrunner,zhou2021high}. As opposed to these works, our work studies a novel attack on lending protocols by examining the CRV short-selling attempt that was facilitated by through lending activity on Aave.

\section{Data collection}
We collect both Ethereum blockchain data, as well as data from centralized exchanges. To gather the on-chain data we run an erigon~\cite{2022erigon} archive node. In particular, we filter the transaction logs for those that relate to the Aave lending market and query the historical state of the Aave contracts to obtain the relevant data. Additionally, we inspect the logs to identify the value and target of all CRV transfers originating from Avi Eisenberg's wallet \cite{Avi_wallet}. For centralized exchanges, we collect data from Binance and OKX. The former, as it is the largest crypto exchange and its price is therefore very reliable. The latter, as the short-seller transferred the vast majority of the borrowed funds to his OKX account. For Binance data, we connect to their API to get aggregated price, and volume data of CRV-BUSD \cite{Binance_API}. OKX provides aggregated trade data on a daily basis. We downloaded all aggregated trade data for the month of November and analyze the four pools they offer containing CRV: Ether (ETH), Bitcoin (BTC), and the two stablecoins USDT and USDC.

\section{Avi Eisenberg's Attack}
\label{sec: Results}

In this section, we focus on Eisenberg's attack. We first discuss an attack he had previously suggested and transition to his actual attack. We trace his money flows to decentralized and centralized exchanges. Then, we investigate why Eisenberg likely selected CRV and what his intentions may have been.

Naturally, lending protocols are required to closely monitor the price of the assets supported on the protocol to determine quantities such as the value of the collateral, line of credit, or the liquidation threshold. An attacker or an external event causing a rapid price change can catch other users of the protocol off guard. For example, a rapid increase in price of a borrowed token or pledged collateral can lead to a short-squeeze, and, in dramatic cases, lead to a situation where not all debt can be retrieved (bad debt). This bad debt is from the perspective of the liquidity providers lost money. Thus, rapid price movements pose a threat to lending protocols. 

This was highlighted in October 2022 when Avi Eisenberg, at the time already well-known for his attack on Mango Markets, began floating the idea of attacking Aave in a series of tweets~\cite{2023avieisenattackdiscussion}. Essentially his idea boils down to using borrowed assets to artificially inflate the value of the pledged collateral, and, in the end, taking out more loans than the original value of the collateral. We discuss his proposed attack in more detail in Appendix~\ref{app_sec:Attack_AVI}. It is important to note that what Eisenberg proposes is in fact an attack on the lending protocol. He has no intention of repaying the debt but solely attempts to borrow more than he posted as collateral. Shortly after his comments and possibly in response, Aave froze the lending pool for REN -- the asset Eisenberg suggested for the attack -- and a few other assets~\cite{2022AAVE-REN-freeze}. 

\begin{figure}[h]
    \centering
    \includegraphics[scale=1]{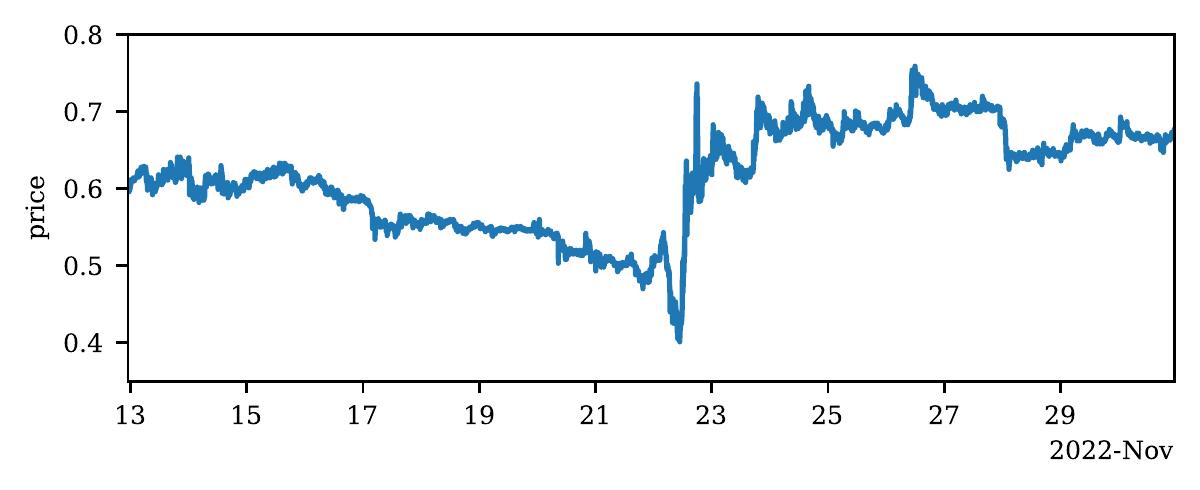}\vspace{-4pt}
    \caption{Price of CRV on Binance during the in November (CRV-BUSD pool). We observe large price movements around November 22, the main date during Eisenberg's attack.}
    \label{fig:CRVpricebinance_nov}\vspace{-8pt}
\end{figure}

In November 2022, Avi Eisenberg himself began targeting Aave by borrowing CRV and dumping it on the market. In Figure~\ref{fig:CRVpricebinance_nov}, we show the CRV price on Binance. We observe a significant and rapid price drop on November 22, which, as discussed below, is also the date Eisenberg dumps most his borrowed CRV on to the market. 

Eisenberg started his move on November 13 by depositing approximately 39 Mio USDC on Aave V2. The next day he took out his first CRV loan. Initially, the short-seller's activity was a relatively small scale. Until the end of November 15, Eisenberg had borrowed 6 Mio CRV (approx 3.6 Mio USDC) of which he sold about 2.2 Mio through the decentralized exchange aggregator 1inch~\cite{20231inch}.  

From November 17 to 22 he stepped up his strategy by further borrowing more than 68 Mio CRV. Eisenberg borrowed nearly all available CRV which in turn also led to a spike in borrowing rates. We show the available liquidity and borrowing rates in Appendix~\ref{app_sec:borrowingrates}. We note that even though the high borrowing rates is an expense the attacker must shoulder, the short duration of the attack hardly makes an annualized borrowing rate of around $300\%$ prohibitive. 

From November 16 on, Avi Eisenberg no longer sold his tokens on 1inch. Rather, he transferred large amounts of borrowed CRV to the centralized exchange OKX -- 71.6 Mio CRV of the total 77 Mio he borrowed over the course of November. While centralized exchanges do not offer the same transparency and it is therefore not possible to exactly track his transactions, we can gather several clues. OKX provides aggregated trade data, i.e., for every trade we know the price at which the trade occurred, the timestamp as well as the transaction size.  

\begin{figure}[h]
    \centering
    \includegraphics[scale=1]{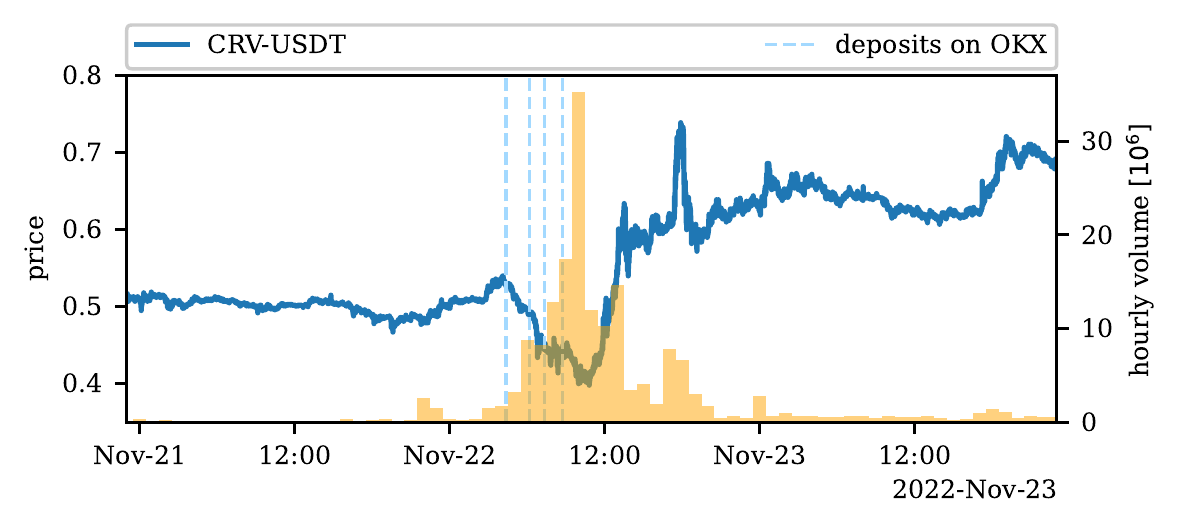} \vspace{-6pt}
    \caption{Plot of the CRV-USDT price on OKX (blue) and the hourly trading volume measured in CRV (yellow). The dashed vertical lines indicate the times at which the short-seller deposited significant amounts of CRV on OKX. We observe significant volume shortly after the these deposits.}
    \label{fig:CRV-USDT_price_OKX_closeup} \vspace{-4pt}
\end{figure}

When analyzing the four pools that OKX offers, CRV-USDT, CRV-UST, CRV-ETH, and CRV-BTC, we notice a conspicuous jump in trading volume in the CRV-USDT pool. In Figure~\ref{fig:CRV-USDT_price_OKX_closeup}, we show the CRV-USDT price and the hourly volume. We observe that the trading volume dramatically jumps after Eisenberg deposited significant funds to OKX -- indicated by dashed vertical lines. In fact, trading volume is orders of magnitudes larger than usual for this pool. Note that while he had already deposited some of the borrowed CRV on OKX before November 22, those amounted to significantly less than the deposits on the 22nd (11.6 Mio CRV before November 22 vs. 60 Mio CRV on November 22).

Thus, it is highly likely that Avi Eisenberg sold his borrowed CRV on OKX primarily for the stablecoin USDT in the hope of causing a large enough price drop. We note that he likely chose a centralized exchange like OKX rather than a decentralized exchange as both Aave's price oracle, Chainlink, as well as arbitrage bots often assume the centralized exchange price to be the `true' price. Further, the liquidity depth is on centralized markets is typically larger and, thus, could likely expect a better price.

Naturally, the question arises why the short-seller ultimately targeted CRV. A lot of the public discourse around Eisenberg's attack focuses on the debt position of the Curve founder, Michael Egorov, who had deposited significant amounts of CRV on Aave and borrowed against it. The theory is that Eisenberg attempted to cause such a price drop, leading to the liquidation of Egorov's debt~\cite{2023Eigenphi}. This liquidation would then lead to a further decrease in price, benefiting the short-seller. We plot the health factor of the Curve founder's debt position in Appendix~\ref{app_sec:Curve Founder's Position on AAVE} in Figure~\ref{fig:health-factor-CRV-founder}. We see that his health factor is around 1.5 around the attack but does fluctuate significantly. 

While it is possible that Eisenberg attempted to get the debt position liquidated, and he did allude to this~\cite{2022EisenbergTweetCRV}, this represents quite a fanciful plan. Egorov's debt was quite far from liquidation, making a large price drop necessary. Additionally, the Curve founder has significant funds at his disposal to post additional collateral -- something he ultimately did. Thus, it appears that this plan would only had slim chances of success to begin with.

Rather than solely focusing on the Curve founder's debt, we stress that there were further reasons for Eisenberg targeting CRV. For an attack relying on borrowed funds to move the price, the amount of liquidity available to borrow must be significant relative to the total market capitalization of the token. Further, if the total market capitalization is too large, the short-seller may simply lack the funds for the attack, even if a large proportion of the assets were available on Aave.

\begin{figure}[h]
\centering
  \begin{subfigure}{0.49\linewidth}
  \centering
    \includegraphics[scale=1]{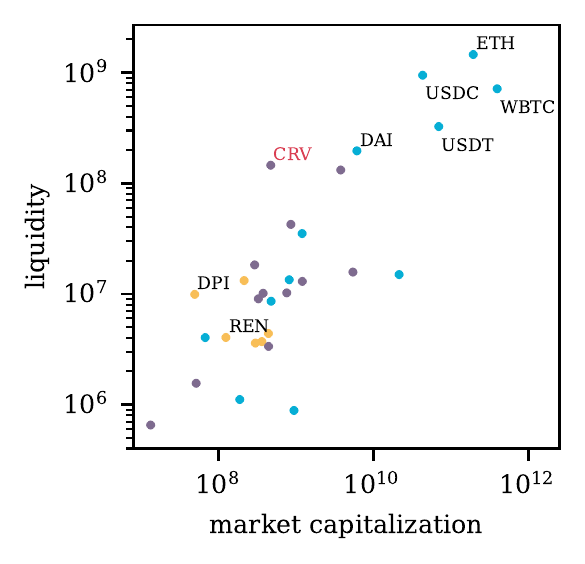} \vspace{-15pt}
    \caption{Liquidity deposited on Aave V2 relative to the token's market cap.}
    \label{fig:marketcap}
  \end{subfigure}
    \hfill
  \begin{subfigure}{0.49\linewidth}
  \centering
    \includegraphics[scale=1]{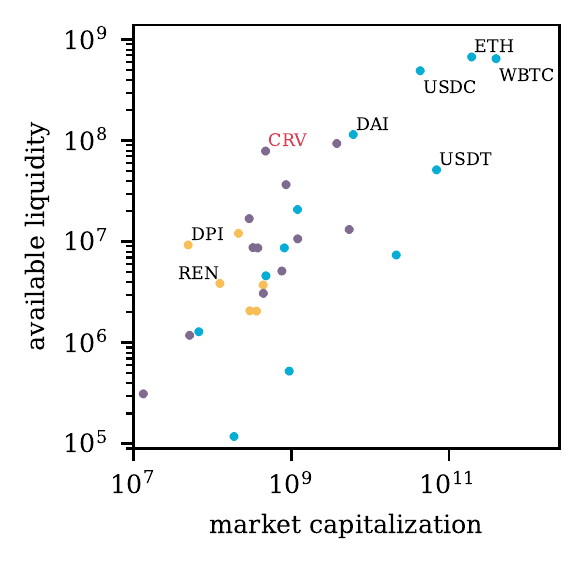} \vspace{-15pt}
    \caption{Liquidity available to borrow on Aave V2 relative to the token's market cap.}
    \label{fig:marketcap1}
\end{subfigure}
  \caption{Plot showing the liquidity (left) and available liquidity (right) of different tokens offered on Aave V2 on November 1 2022 as a function of market capitalization. For the short seller CRV is an attractive choice as it has a large proportion of its market cap on Aave (top left corner in the respective plots). Tokens for which Aave froze lending before Eisenberg's attack are represented as yellow points. Purple points represent tokens for which lending was de-activetad after the attack and for tokens plotted in blue borrowing remains possible.}
  \label{fig:liquidity_vs_marketcap} \vspace{-0pt}
\end{figure}

In Figure~\ref{fig:marketcap}, we plot the liquidity deposited as a function of the tokens total market capitalization for each asset available on Aave V2. The data is as of November 1 and the market capitalization data is taken from CoinMarketCap. 
Similarly, Figure~\ref{fig:marketcap1} shows the liquidity \textit{available} to a borrower as a function of the market capitalization. The color of the data points has the following meaning: we plot tokens for which Aave disabled borrowing before Eisenberg's attack but after his October tweets in yellow. Purple points represent tokens for which Aave froze borrowing after the attack (such as CRV) and assets that can still be borrowed are displayed in blue.

As we can see, CRV must have seemed a very attractive option for Eisenberg given that the available liquidity relative to the market capitalization was very high at more than 15\% (top left corner of the plot). We conclude from Figure~\ref{fig:liquidity_vs_marketcap} that CRV was Eisenberg's best option for the attack. 

This highlights that CRV was an attractive option for Eisenberg irrespective of the debt of Egorov. We further point out that Eisenberg could have also been attempting a modification of his originally proposed attack discussed in Appendix~\ref{app_sec:Attack_AVI}. 
\begin{enumerate}
    \item The attacker deposits USDC to borrow significant amounts of CRV
    \item The attacker sells all CRV on a centralized exchange lowering the price. The funds received cover some of the debt taken out but also lead to a devaluation of the debt and a reduction in debt to collateral ratio.
    \item The attacker uses the lower debt level to take out a loan in USDT. 
\end{enumerate}
The proceeds in step 2) as well as the loan in step 3) would fund his attack. A liquidation of Egorov's debt or panic-selling of CRV holders during the price devaluation would boost the attacker's profit. Note that strategy should be considered an attack rather than short-selling, as unlike a short-seller the attacker has no intentions of repaying the borrowed tokens. This underscores that the ramifications of such strategies for lending protocols are far greater than mere short-selling.

\section{Aftermath}

\begin{figure}
    \centering
    \includegraphics[scale=1]{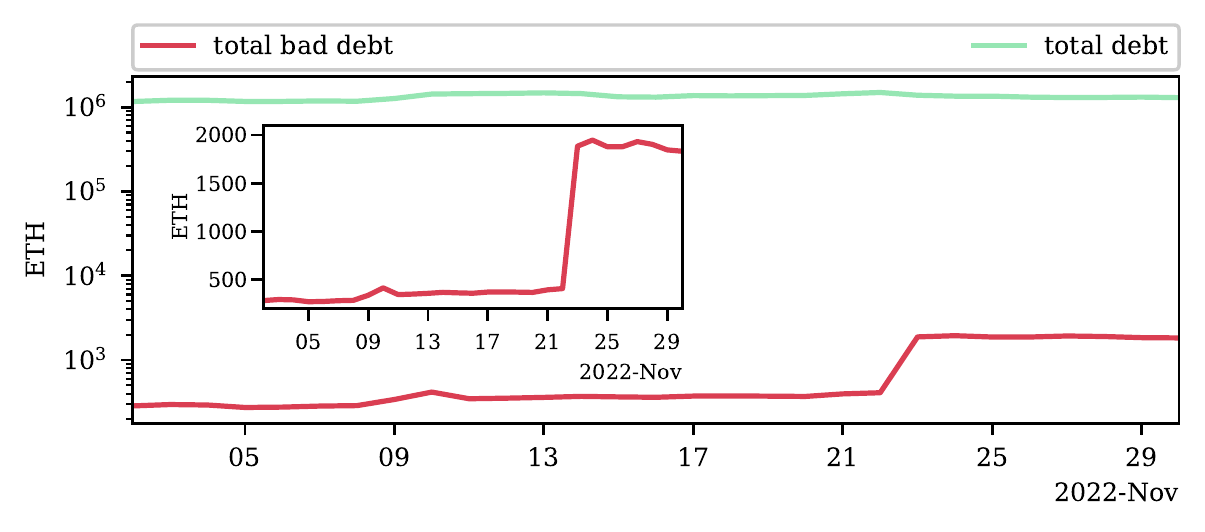} \vspace{-15pt}
    \caption{Plot of the total debt (green) and bad debt, i.e., the debt of liquidated positions that has become irretrievable, on Aave V2. Note that the short-squeeze of Avi Eisenberg quadrupled the amount of bad debt, highlighting the strain the actions of the short-seller exerted on the protocol.\vspace{-4pt}}
    \label{fig:bad_debt_aave:}
\end{figure}

After initially managing to reduce the price of CRV by around $20\%$, the CRV price surged later on November 22, jumping by about $75\%$ (cf. Figure~\ref{fig:CRV-USDT_price_OKX_closeup}). This rapid price jump led to the liquidation of Eisenberg's debt. However, the sheer size of his debt had negative implications for the protocol. In Figure~\ref{fig:bad_debt_aave:}, we plot the total value of all bad debt as well as all debt on Aave V2 measured in ETH. Note that we obtain the total (bad) debt by identifying all users that ever borrowed assets on Aave V2. We then, on a daily basis, query the value of their collateral and debt. The debt of positions with a debt value larger than the collateral value is bad debt. We observe that the total amount of bad debt more than quadruples due to the activities of Eisenberg. This highlights the severe impact that the short-seller's activity had. Furthermore, we note that this is the damage incurred due to the failed attack. Had the attack succeeded the implications would have likely been far graver. Furthermore, we note that in this instance the bad debt was, in the end, covered by the protocol.

It has been suggested that the delay of the price oracle has aggravated the bad debt. In Appendix~\ref{app_sec:Latency-of-Price-Oracle}, we plot the health factor of Avi Eisenberg both using the Binance price as well as the oracle price. While differences are visible, the ultimate effect of this price delay is rather limited.

What, however, could cause potentially fatal delays is Aave's reliance on comparatively slow moving DAO votes. While the protocol has a proven track-record of successfully updating its risk parameters to face changing market environment such as during the Merge~\cite{heimbach2023defi}, these changes do require a few days. In this instance, Aave was not oblivious to the threats of Eisenberg's proposed attack and froze borrowing of assets like REN (yellow points in Figure~\ref{fig:liquidity_vs_marketcap}) only weeks after Eisenberg mused about this potential attack~\cite{2022AAVE-REN-freeze}. While we cannot be certain that Aave disabled borrowing for these tokens as a consequence of Eisenberg's comments, the timing of their actions and the discussions in the governance forum lead us to believe that the two are not unrelated. However, reacting to CRV developments proved difficult as the majority of the short-selling occurred on a single day. In the aftermath of the attack Aave V2 disabled several further tokens including CRV (purple points in Figure~\ref{fig:liquidity_vs_marketcap})~\cite{2022AAVE-CRV-freeze}. 

\section{Discussion}
\label{sec:Discussion}

Given their size lending protocols like Aave are rightly considered one of the most successful applications of DeFi. However, the attack of Avi Eisenberg demonstrated that despite -- or in some cases due to -- this success vulnerabilities remain. An attack like the one proposed by Eisenberg is facilitated by two key features.

First, Eisenberg relied on Aave holding a significant portion of all available CRV tokens ($>30\%$ of market cap). The large amount of funds available to borrow enabled the attacker to significantly move the price of the asset, thus, effectively manipulating the price and thereby altering the valuation of all CRV debt and collateral. This avenue of attack is only open due to Aave V2's success in attracting such a significant portion of total liquidity. 

Second, Aave V2 relies on a set of risk parameters that can only be altered by a relatively slow DAO vote. Thus, during such an attack the parameters are, in effect, static, hindering any response from the protocol during the attack. 

To face this challenge lending protocols have multiple options. One, they can restrict themselves to large market cap tokens whose price is far harder to manipulate. This is a route Aave V2 has pursued by freezing lending of several assets like CRV. Two, they could make their parameters more restrictive by requiring more collateral or limiting the amount of available funds relative to the respective token's market capitalization. However, all these options limit the scope of the protocol and may make the protocol less attractive to certain users.

Alternatively, they can rely more heavily on active risk management, allowing the risk parameters to change rapidly in response to altering market conditions. As has been suggested, this could be accomplished by employing a feedback loop that automatically updates the risk parameters~\cite{2022report-bad-debt}. This would, however, inadvertently add additional layers of complexity and potentially open new avenues of attack as the feedback loops would likely rely on further market parameters fed by oracles.

Instead, lending protocols could rely on an active risk manager who has the flexibility to intervene. This is the route Aave has chosen for its newest version, V3. On Aave V3, the DAO can elect a `risk admin' who has the power to change the risk parameters without a governance vote~\cite{2023AAVEV3risk}. However, this naturally compromises the protocol's decentralization.

Finally, we note that a future attack could use multiple protocols to borrow a total amount sufficient for price manipulation. Thus, with the growing popularity of DeFi lending protocols, the inter-dependencies and complexities will also increase, necessitating cross-protocol risk management.

\section{Conclusion}

In this work, we use Avi Eisenberg's actions as a case study. We investigate his money flows across both DeFi and centralized exchanges. Furthermore, we show that Eisenberg selected CRV as, among other reasons, Aave at the time had more than $30\%$ of the market capitalization locked on its protocol, making large-scale price manipulations feasible. Despite the failure of the attack, it still led to irretrievable debt worth more than 1.5 Mio USD at the time, quadrupling the protocol’s irretrievable debt. 

In summary, Eisenberg's attack highlighted that DeFi lending protocols that contain a large portion of a token's market capitalization enable price manipulations. The slow-moving governance votes are ill-suited to react to such an attack. This has left large lending protocols in predicament: either limit their offering or place their trust in active risk management.

%
%
\bibliographystyle{splncs04}
\bibliography{references}

\begin{thebibliography}{10}
\providecommand{\url}[1]{\texttt{#1}}
\providecommand{\urlprefix}{URL }
\providecommand{\doi}[1]{https://doi.org/#1}

\bibitem{2021aave}
Aave. \url{https://aave.com/} (2022)

\bibitem{2022EisenbergTweetCRV}
Eisenberg {Tweet} from {Nov-20}.
  \url{https://twitter.com/avi\_eisen/status/1594293743
  380615168?cxt=HHwWgMCinc7tiKAsAAAA} (2022)

\bibitem{2022report-bad-debt}
Report on code at risk: {An} {In-depth} {Analysis} of {How} {AAVE's} \$1.6
  {Million} {Bad} {Debt} {Was} {Created}.
  \url{https://drive.google.com/file/d/1u3vtcsQ1qfclt6Od8aZx5DkvuQRE4GMH/view}
  (2022)

\bibitem{2022AAVE-REN-freeze}
Risk {Parameter} {Updates} for {Aave} {V2} {ETH} {Market} (2022-11-12).
  \url{https://app.aave.com/governance/proposal/?proposalId=117} (2022)

\bibitem{2022AAVE-CRV-freeze}
Risk {Parameter} {Updates} for {Aave} {V2} {Ethereum} {Liquidity} {Pool}
  (2022-11-25).
  \url{https://governance.aave.com/t/risk-parameter-updates-for-aave-v2-ethereum-liquidity-pool-2022-11-25/10824}
  (2022)

\bibitem{20231inch}
1inch. \url{https://app.1inch.io} (2023)

\bibitem{2023AAVEV3risk}
Aave {V3} {Risk}. \url{https://docs.aave.com/risk/} (2023)

\bibitem{Binance_API}
Binance {API}. \url{https://github.com/binance/binance-public-data} (2023)

\bibitem{2023chainlink}
Chainlink. \url{https://chain.link/} (2023)

\bibitem{2023avieisenattackdiscussion}
Chainlink. \url{https://twitter.com/avi\_eisen/status/1582763707742183424}
  (2023)

\bibitem{2023curve}
Curve. \url{https://curve.fi/} (2023)

\bibitem{Mango_Markets}
Mango {Markets}. \url{https://mango.markets/} (2023)

\bibitem{blockworks_Avi_attack_Mango}
Mango {Markets} {Mangled} by {Oracle} {Manipulation} for 112{M}.
  \url{https://blockworks.co/news/mango-markets-mangled-by-oracle-manipulation-for-112m}
  (2023)

\bibitem{Avi_wallet}
Wallet of {Avi} {Eisenberg}: 0x57e04786e231af3343562c062e0d058f25dace9e (2023)

\bibitem{CRV_wallet}
Wallet of {Curve} {Founder}: 0x7a16ff8270133f063aab6c9977183d9e72835428 (2023)

\bibitem{aramonte2022defi}
Aramonte, S., Doerr, S., Huang, W., Schrimpf, A., et~al.: Defi lending:
  intermediation without information? Tech. rep., Bank for International
  Settlements (2022)

\bibitem{bartoletti2021sok}
Bartoletti, M., Chiang, J.H.y., Lafuente, A.L.: Sok: Lending pools in
  decentralized finance. In: International Conference on Financial Cryptography
  and Data Security (2021)

\bibitem{chiu2022inherent}
Chiu, J., Ozdenoren, E., Yuan, K., Zhang, S.: On the inherent fragility of defi
  lending  (2022)

\bibitem{2023Eigenphi}
Eigenphi: Cleaning {Up} of the {Battlefield} of {Avi} and {Curve}.
  \url{https://eigenphi.substack.com/p/cleaning-up-of-the-battlefield-of}
  (2022)

\bibitem{eskandari2021sok}
Eskandari, S., Salehi, M., Gu, W.C., Clark, J.: Sok: Oracles from the ground
  truth to market manipulation. arXiv preprint arXiv:2106.00667  (2021)

\bibitem{gudgeon2020defi}
Gudgeon, L., Werner, S., Perez, D., Knottenbelt, W.J.: Defi protocols for
  loanable funds: Interest rates, liquidity and market efficiency. In:
  Proceedings of the 2nd ACM Conference on Advances in Financial Technologies.
  pp. 92--112 (2020)

\bibitem{heimbach2023defi}
Heimbach, L., Schertenleib, E., Wattenhofer, R.: Defi lending during the merge.
  arXiv preprint arXiv:2303.08748  (2023)

\bibitem{2022erigon}
{ledgerwatch}: Erigon. \url{https://github.com/ledgerwatch/erigon} (2023)

\bibitem{perez2021liquidations}
Perez, D., Werner, S.M., Xu, J., Livshits, B.: Liquidations: Defi on a
  knife-edge. In: International Conference on Financial Cryptography and Data
  Security. pp. 457--476. Springer (2021)

\bibitem{qin2021empirical}
Qin, K., Zhou, L., Gamito, P., Jovanovic, P., Gervais, A.: An empirical study
  of defi liquidations: Incentives, risks, and instabilities. p. 336–350. IMC
  '21 (2021)

\bibitem{qin2022quantifying}
Qin, K., Zhou, L., Gervais, A.: Quantifying blockchain extractable value: How
  dark is the forest? In: 2022 IEEE Symposium on Security and Privacy (SP). pp.
  198--214. IEEE (2022)

\bibitem{torres2021frontrunner}
Torres, C.F., Camino, R., State, R.: Frontrunner jones and the raiders of the
  dark forest: An empirical study of frontrunning on the ethereum blockchain.
  arXiv preprint arXiv:2102.03347  (2021)

\bibitem{yaish2022blockchain}
Yaish, A., Tochner, S., Zohar, A.: Blockchain stretching \& squeezing:
  Manipulating time for your best interest. In: Proceedings of the 23rd ACM
  Conference on Economics and Computation. pp. 65--88 (2022)

\bibitem{zhou2021high}
Zhou, L., Qin, K., Torres, C.F., Le, D.V., Gervais, A.: High-frequency trading
  on decentralized on-chain exchanges. In: 2021 IEEE Symposium on Security and
  Privacy (SP). pp. 428--445. IEEE (2021)

\end{thebibliography}
\appendix

\section{Borrowing Rates}
\label{app_sec:borrowingrates}

The parameters of Aave's smart contract determine the terms of borrowing and lending. The rates are ultimately determined by the \textit{utilization}, i.e., the fraction of deposited funds that have been borrowed. In the case of CRV the borrowing rate at time $t$ as a function of the utilization $U_t$ is given as
\begin{equation*}
r_t = \begin{cases}
        r_0 + \frac{U_t}{U_{\text{optimal}}} r_{\text{slope}_1} & \text{if }U_t \leq U_{\text{optimal}},  \\
        r_0 + r_{\text{slope}_1} +\frac{U_t-U_{\text{optimal}}}{1-U_{\text{optimal}}} r_{\text{slope}_2} & \text{if }U_t > U_{\text{optimal}}, \\
       \end{cases}
\end{equation*}
with $r_0 = 0$, $U_{\text{optimal}} = 45\%$, $r_{\text{slope}_1} = 7\%$, and $r_{\text{slope}_2}=300\%$. Note that the rate $r_t$ is given as an annualized rate and has a kink at the `optimal utilization' $U_{\text{optimal}}$. Furthermore, the rates are capped at a maximum of $r_{\text{slope}_1}+r_{\text{slope}_2}=307\%$. The aforementioned risk parameters differ across cryptocurrencies and are set by the protocol in accordance to the perceived risk of the asset. 

\begin{figure}[h]
\centering
  \begin{subfigure}{0.48\linewidth}
  \centering
    \includegraphics[scale=1]{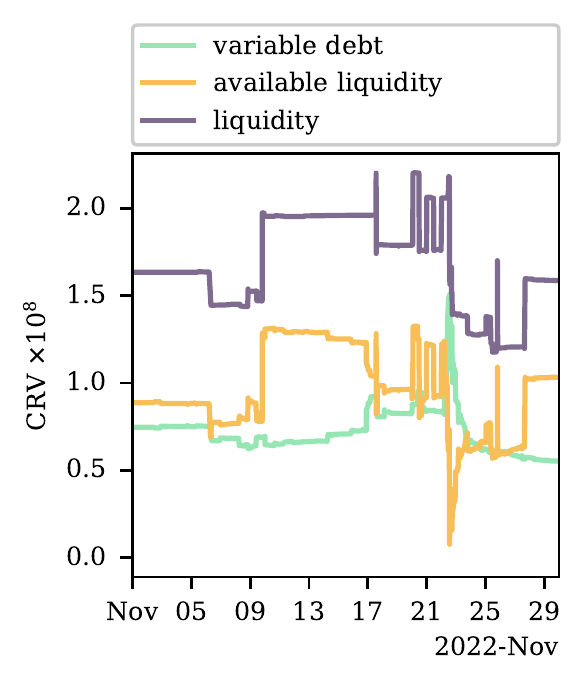}\vspace{-6pt}
    \caption{Debt (green), available liquidity (yellow), and total liquidity (violet) in Aave's CRV pool.}
    \label{fig:CRV_liquidity_aave}
  \end{subfigure}
    \hfill
  \begin{subfigure}{0.48\linewidth}
  \centering
    \includegraphics[scale=1]{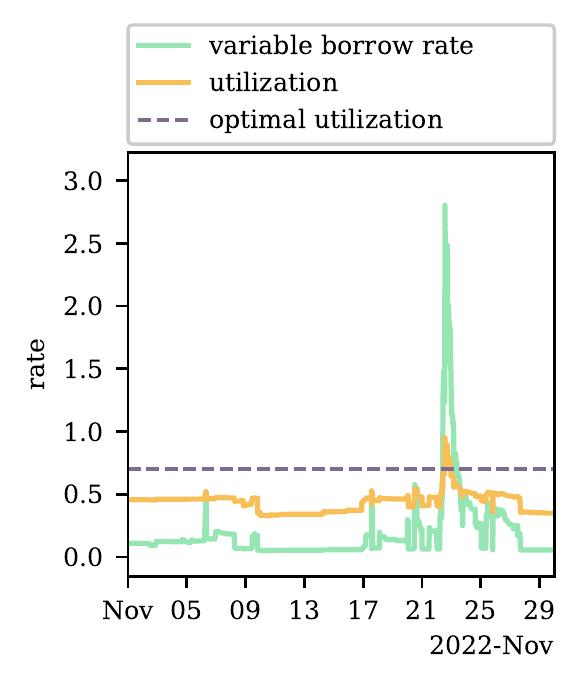}\vspace{-6pt}
    \caption{Borrowing rate (green) and utilization (yellow) of CRV on Aave V2 during November.}
    \label{fig:CRV_rates_aave_nov}
\end{subfigure}
  \caption{Total volume of debt taken out (left) and borrowing rates (right) are both ploted in green. Note the total debt taken out peaked at about 150 Mio CRV (approx. 90 Mio USD at the time) which corresponds to basically all liquidity that was available on Aave V2 at the time. } 
\end{figure}

Eisenberg started his move on Nov-13 by depositing approximately 39 Mio USDC on Aave V2. The next day he took out his first CRV-loan. Initially, the short-seller's activity was on quite a small scale. Until the end of November 15, Eisenberg had borrowed 6 Mio CRV (approx 3.6 Mio USDC) of which he sold about 2.2 Mio through the decentralized exchange 1inch~\cite{20231inch}. From November 17 to 22 he stepped up his strategy by borrowing more than 68 Mio CRV. As shown in Figure~\ref{fig:CRV_liquidity_aave} the available liquidity (in yellow) in Aave's CRV-pool dropped to zero, signaling that the short-seller borrowed the maximal possible amount. Consequently, the borrowing rates also spiked (cf. Figure~\ref{fig:CRV_rates_aave_nov}).

\section{Latency of Price Oracle}
\label{app_sec:Latency-of-Price-Oracle}
\begin{figure}
    \centering
    \includegraphics[scale=1]{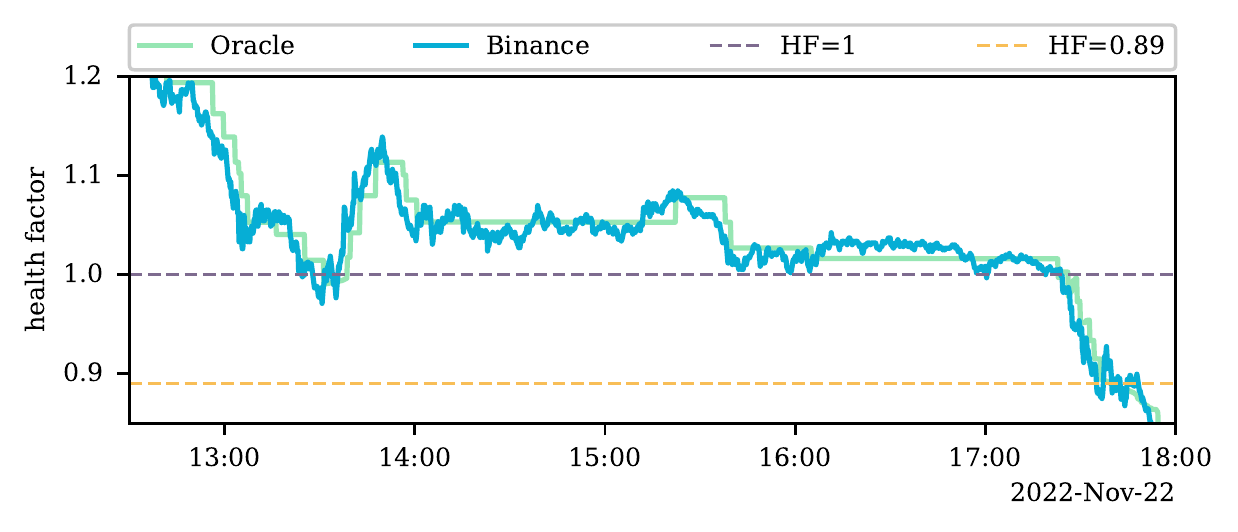}
    \caption{Health factor of Eisenberg using the oracle price (green) and the Binance price (blue). A health factor below 1 makes liquidations possible, whereas, the yellow line indicates the price at which bed debt would remain.}
    \label{fig:app_CRVhealth}
\end{figure}

Apart from the danger's posed by rapid price changes, inaccuracies of the protocol's oracle price can also pose significant threats. In Figure~\ref{fig:app_CRVhealth}, we plot the health factor of Eisenberg using both Aave's oracle price as well as the Binance price. While small differences are visible the oracle price reflected the Binance quite accurately. Observe that between 5 and 6 p.m. the price increase rapidly bringing down the health factor from just above 1 to under 0.89 in a matter of minutes, the later threshold indicates the value at which irretrievable debt begins to accumulate. This highlights that the time window for liquidations can be very short and was very short for this case study. The liquidations, nevertheless, succeeded in retrieving the majority of the debt. However, the combination of the sheer volume of the position's debt and the limited time the protocol had to liquidate the position, in the end, left the protocol with the significant amount of irretrievable debt which in the end the protocol covered.

\section{Curve Founder's Position on Aave}
\label{app_sec:Curve Founder's Position on AAVE}
\begin{figure}[t]
    \centering
    \includegraphics[scale=1]{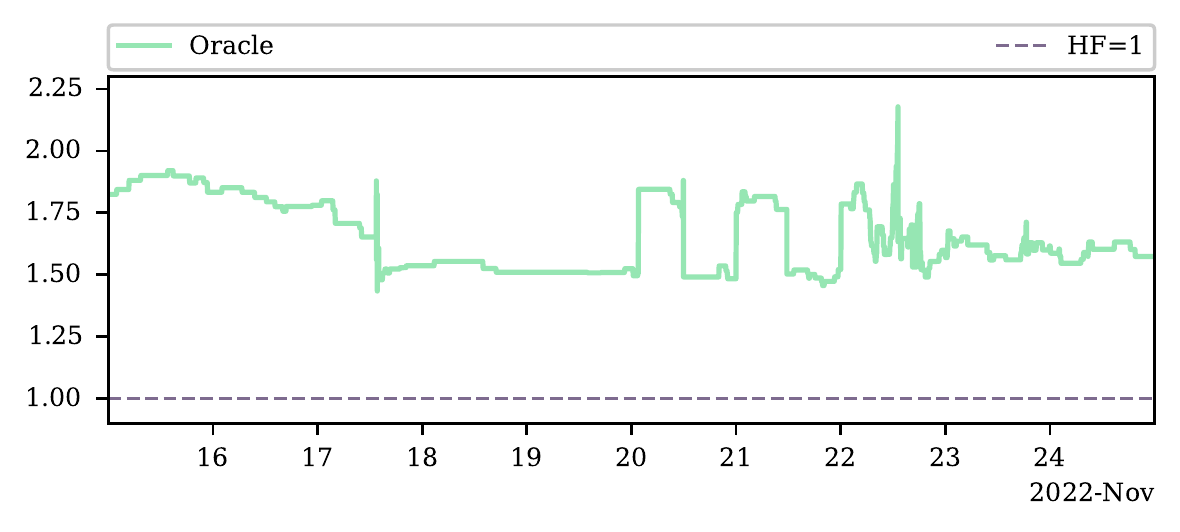}
    \caption{Health factor of Curve founder using the oracle price (green). A health factor below 1 makes liquidations possible, the position did not come close to the liquidation threshold.}
    \label{fig:health-factor-CRV-founder}
\end{figure}

In Figure~\ref{fig:health-factor-CRV-founder} we plot the health factor of the position of the Curve founder's wallet on Aave~\cite{CRV_wallet}. Even though, it was suggested by Avi Eisenberg himself and repeated in multiple articles discussing the attack, we do not believe that Avi Eisenberg attack was solely focused around causing a liquidation of this position. While liquidations of a position with such a significant volume of CRV collateral would have lead to a further decrease in price, we believe that this strategy would have been fanciful. The position's health factor never dropped significantly below a health factor of 1.5 during the attack and it is further hard to believe that the Curve founder would not have the funds available to prevent his liquidation.

\section{Attack proposed by Avi Eisenberg}
\label{app_sec:Attack_AVI}

Avi Eisenberg suggested the following attack on Aave's REN-pool~\cite{2023avieisenattackdiscussion}. An attacker has two wallets, A and B, where initially $100\%$ of his assets are in wallet A. We here sketch the idea of the attack:
\begin{enumerate}
    \item From wallet A deposit $100\%$ of the assets as USDC as collateral to borrow all REN possible on Aave. Using USDC as collateral, a user can borrow $85\%$ of the USDC value (loan to value). 
    \item The attacker's borrowed REN ($85\%$ of initial assets) is transferred to wallet B. The attacker uses the REN as collateral to borrow USDC on Aave. At the time the loan to value for REN was $60\%$, so from wallet B the attacker could borrow approximately $85\%\cdot 60\% \approx 50\%$ of the initially deposited USDC. 
    \item The attacker uses the borrowed USDC to buy REN on centralized exchanges to drive up the price of REN\footnote{The thereby acquired REN could be used to go back to step 2 to borrow more USDC to repeat the cycle. Given the loan to value of $60\%$, the hypothetical limit the $50\%$ of assets borrowed in step 2 can be increased to using this loop is given by the geometric series, i.e. $50\%/(1-60\%)= 125\%$.}.
    \item This increases the value of the collateral of wallet B, allowing the attacker to take out more USDC loans. 
\end{enumerate}
The collateral in wallet A will be liquidated by the price increase caused in step 3. Additionally, given the price increase was created by artificial buying pressure, it is likely that the price of REN reverts close to it's original value, thus, also liquidating the collateral in step 2. However, if this price increase is sufficiently large, the additional USDC that can be extracted in step 4 as well as to proceeds of selling the REN acquired in step 3, pays for the attack. 

Note that an attacker may deliberately target a lending protocol for more reasons than pure profits. An attacker looking to damage a protocol, thus, might execute this attack for a small profit or even at a small cost just to make a point.

Finally, we note that for this attack to work, the amount of REN available on Aave must be significant to allow the attacker to use the borrowed REN to create a significant uptick in price. Thus, the proportion of the market cap available on Aave for borrowing must be significant. 

\end{document}